\documentclass[prl, twocolumn, superscriptaddress]{revtex4-1}
\usepackage{amsmath}
\usepackage[english]{babel}
\usepackage[pdftex]{graphicx}
\usepackage{color}
\usepackage{ulem}
\usepackage{comment}

\usepackage[sort&compress]{natbib}
\usepackage{hyperref}

\graphicspath{{images/}{./}}

\usepackage{amsmath}
\usepackage{amssymb}
\usepackage{mathtools}
\usepackage{braket}

\renewcommand{\vec}[1]{\ensuremath{\boldsymbol{#1}}}

\newcommand{\sgn}{\,\mbox{\rm sgn}}

\begin{document}
\title{Antiresonances of Wannier-Stark ladders in Su-Schrieffer-Heeger lattices}
\author{Yonatan Betancur-Ocampo}
\email{ybetancur@fisica.unam.mx}
\affiliation{Instituto de F\'isica, Universidad Nacional Aut\'onoma de M\'exico, Ciudad de México, Mexico}

\author{Guillermo Monsivais}
\email{monsi@fisica.unam.mx}
\affiliation{Instituto de F\'isica, Universidad Nacional Aut\'onoma de M\'exico, Ciudad de México, Mexico}

\begin{abstract}
    We demonstrate that Wannier-Stark ladders (WSLs) in one-dimensional Su-Schrieffer-Heeger (SSH) chains manifest as antiresonance signatures in the electron transmission.  We study the quantum transport in the SSH chain in the presence of a periodic potential profile plus a uniform electric field. By employing the global matrix method with non-trivial matching conditions, we find that these WSL antiresonances correspond to states that are exponentially localized at the potential's incident boundary. These findings provide a robust, unified framework for identifying WSLs across diverse scales, ranging from electronic transport in trans-polyacetylene to classical-wave analogs in phononic crystals and surface water waves.
\end{abstract} 

\maketitle

\section{Introduction}

The study of periodic potentials subject to external biasing fields has long been a cornerstone of condensed matter physics. For a system defined by the potential $V(x) = V_p(x) + Fex$, where $V_p(x)$ is a periodic potential of period $p$ and $F$ represents a constant electric field acting on a charge $e$, the resulting eigenspectrum undergoes a fundamental transformation. In the absence of the field ($F=0$), the spectrum is characterized by traditional energy bands and gaps. However, the introduction of the linear field term $Fex$ breaks the translational symmetry, giving rise to Wannier-Stark Ladder Resonances (WSLR) \cite{Wannier1960,Bloch1929,DiCarlo1994,Helm1999,Zhang2025}. These resonances appear as discrete energy levels inside the continuum spectrum, which are equally spaced by a constant $\Delta E$, forming the so-called Wannier-Stark ladder.

This phenomenon is intrinsically linked to Wannier-Stark Localization (WSLoc), where the wave function of each resonance localizes within a distinct spatial zone \cite{Guo2021,Voisin1988,Emin1987,Bhakuni2019,Monsivais2003,RodriguezRamos2004,Kim2020a}. Although the wave functions corresponding to the different rungs of a given Wannier-Stark ladder (WSL) show a similar profile, they are spatially shifted relative to one another. This framework naturally gives rise to {\it tilted straight bands} \cite{Zener1934,Wannier1960} and hosts both interband Zener tunneling \cite{Wannier1960} and Bloch oscillations \cite{Merlin2024}. Driven by the electric field, these oscillations manifest as the periodic spatial translation of wave packets, which ultimately reflect due to destructive interference within the system.

Although originally formulated in the context of theoretical quantum mechanics, Wannier-Stark physics has been extended to a diverse range of classical, semi-classical, and quantum platforms. Experimental realizations span a variety of wave-based systems, including surface acoustic waves \cite{Lima2010}, coupled electromagnetic cavities \cite{Mukherjee2015a}, optical lattices \cite{Madison1999}, plasmonic waveguide arrays \cite{Wetter2022}, semiconductor superlattices \cite{Voisin1988,LanzillottiKimura2005}, 1D granular crystals \cite{Shi2018}, and elastic media supporting torsional and bending waves \cite{Gutierrez2006,Monsivais2007}. More recently, investigations have focused on acoustic systems, demonstrating robust Bloch oscillations across a broad frequency spectrum \cite{Hartmann2004}.

Among the various models used to explore topological and transport properties in condensed matter, the Su-Schrieffer-Heeger (SSH) model remains one of the most versatile \cite{Fradkin1983,Li2014,Su1979,Heeger1981,Heeger1988,Obana2019}. Originally proposed to describe topological phase transitions in trans-polyacetylene through the modulation of hopping parameters \cite{Asboth2016}, its low-dimensional architecture represents the simplest framework for exploring topological insulators, electronic structures, and tight-binding dynamics.

Identically to Wannier-Stark ladders, the experimental reach of the SSH and one-dimensional tight-binding models has expanded far beyond condensed matter systems, spanning diverse artificial lattices and platforms. These include semiconductor quantum dots \cite{Kiczynski2022}, superconducting circuits \cite{Splitthoff2024}, cold atoms \cite{Meier2016}, Rydberg atoms \cite{Kanungo2022}, phononic chains \cite{Tang2025,Wang2026}, and polariton lattices \cite{Pernet2022}. Furthermore, extended SSH models, incorporating non-Hermitian chains  \cite{Lieu2018,Fan2022,ManyManda2024}, two-dimensional lattices \cite{Kim2020,Xu2020,Li2022,Chen2022a}, and bearded geometries \cite{CaceresAravena2022,BetancurOcampo2024a,Jakubsky2024}, have uncovered unusual topological properties that significantly advance our understanding of topological physics.

In this work, we propose a model that allows us to identify the signatures of Wannier-Stark ladders in a modified SSH lattice. We perform a numerical analysis of wave propagation through the system. Unlike the standard WSLRs observed in traditional quantum mechanics, we find that for SSH chains, the ladder structure is more clearly manifested in the {\it antiresonance} energies. Our analysis focuses on determining these energies and the corresponding wave functions. We highlight also the phenomenological differences that arise when the governing equation comes beyond the standard Schrödinger equation.

\section{Junctions of Su-Schrieffer-Hegger lattices: Boundary conditions}

\begin{figure}
    \centering
    \includegraphics[width=1\linewidth]{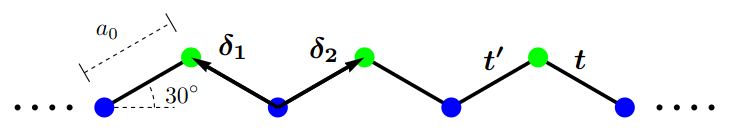}
    \caption{Schematic diagram of the Su-Schrieffer-Heeger (SSH) lattice. Blue and green circles represent the two distinct sublattices, with black lines indicating nearest-neighbor bonds of length $a_0$. The vectors $\vec{\delta}_1 = a_0\left(-\frac{\sqrt{3}}{2}, \frac{1}{2}\right)$ and $\vec{\delta}_2 = a_0\left(\frac{\sqrt{3}}{2}, \frac{1}{2}\right)$ define the relative atomic positions, while $t$ and $t'$ denote the alternating hopping parameters.}
    \label{SSH_latt}
\end{figure}

The Su-Schrieffer-Heeger (SSH) model undergoes a topological phase transition embodied in the Bloch Hamiltonian \cite{Asboth2016}  

\begin{equation}\label{Hssh}
    H_\textrm{SSH}(k) = \left(\begin{array}{cc}
       0  & t-t'\textrm{e}^{-ik} \\
       t-t'\textrm{e}^{ik}  & 0
    \end{array}\right),
\end{equation}

\noindent for the polyacetylene chain shown in Fig. \ref{SSH_latt}, where we have used the displacement $k \rightarrow k + \pi/2$ and the length of the unit cell is $a_c = 1$. The tuning of one of the hopping parameters leads to a trivial phase when $t < t'$ or a topological one for $t > t'$ \cite{Su1979,Heeger1988,Meier2016,Li2018,Lieu2018,Xie2019,Thatcher2022,Kim2020a,Manda2023,BetancurOcampo2024,ManjarrezMontanez2025,Jakubsky2026,DalPoggetto2024}.  The winding number is the topological invariant that characterizes these phases. This transition strongly influences both interband and intraband tunneling of $pnp$ junctions \cite{BetancurOcampo2026}. 

The dispersion relation is obtained by diagonalizing the Bloch Hamiltonian in Eq. \eqref{Hssh}

\begin{equation}
    E_s(k) = s|t-t'\textrm{e}^{-ik}|,
\end{equation}

\noindent where $s = \pm 1$ is the band index. This dispersion relation warrants the reflection law given by $k_\textrm{in} = -k_\textrm{r}$. 

For abrupt changes in the electrostatic potential $V(x)$, the matching conditions are provided by integrating the Schr\"odinger equation into a narrow range $(-\epsilon,\epsilon)$ around the interface. From the Bloch Hamiltonian in Eq. \eqref{Hssh} of the SSH lattice, we get

\begin{eqnarray}\label{ISE}
    \lim_{\epsilon \rightarrow 0}\int\limits^\epsilon_{-\epsilon}\left(\begin{array}{cc}
        V(x) & \hat{g}^*(k) \\
        \hat{g}(k) & V(x)
    \end{array}\right)\left(\begin{array}{c}
   \psi_1(x)\\
   \psi_2(x)\end{array}\right)dx & = & 0
\end{eqnarray}

\noindent The operator $\hat{g}(k)$ is defined by

\begin{eqnarray}\label{gf}
   \hat{g}(k) & = & t-t'\textrm{e}^{ik} \nonumber\\
    & = & t -t'\sum^\infty_{n = 0}\frac{1}{n!}\frac{d^n}{dx^n}
\end{eqnarray}

\noindent and

\begin{equation}
    \hat{g}^*(k)=t -t'\sum^\infty_{n = 0}\frac{(-1)^n}{n!}\frac{d^n}{dx^n},
\end{equation}

\noindent where the linear momentum operator is given by $k = -i\frac{d}{dx}$. Assuming that the wavefunctions $\psi_j(x)$ (for $j = 1, 2$) and the potential $V(x)$ are free of singularities, we take the limit $\epsilon \rightarrow 0$. Substituting the operators $\hat{g}(k)$ and $\hat{g}^*(k)$ into the integrals yields the equations

\begin{equation}\label{int}
    \sum^\infty_{n=0}\frac{(-1)^{(j+1)n}}{n!}\lim_{\epsilon \rightarrow 0}\int\limits^\epsilon_{-\epsilon}\frac{d^n\psi_j(x)}{dx^n}dx  = 0,  
\end{equation}

\noindent which give rise to the matching conditions of wave functions

\begin{eqnarray}\label{TCC}
    \left.\hat{Q}_j\psi_j(x)\right|_{x=0^-} & = &\left.\hat{Q}_j\psi_j(x)\right|_{x=0^+},
\end{eqnarray}

\noindent where the operator $\hat{Q}_j$ is defined as

\begin{equation}\label{Qs}
    \hat{Q}_j =  \sum^\infty_{n = 0}\frac{(-1)^{(j+1)n}}{(n+1)!}\frac{d^n}{dx^n}.
\end{equation}

\begin{figure}
    \centering
    \begin{tabular}{c}
    (a) \qquad \qquad \qquad \qquad \qquad \qquad \qquad \qquad\\
    \includegraphics[width=1\linewidth]{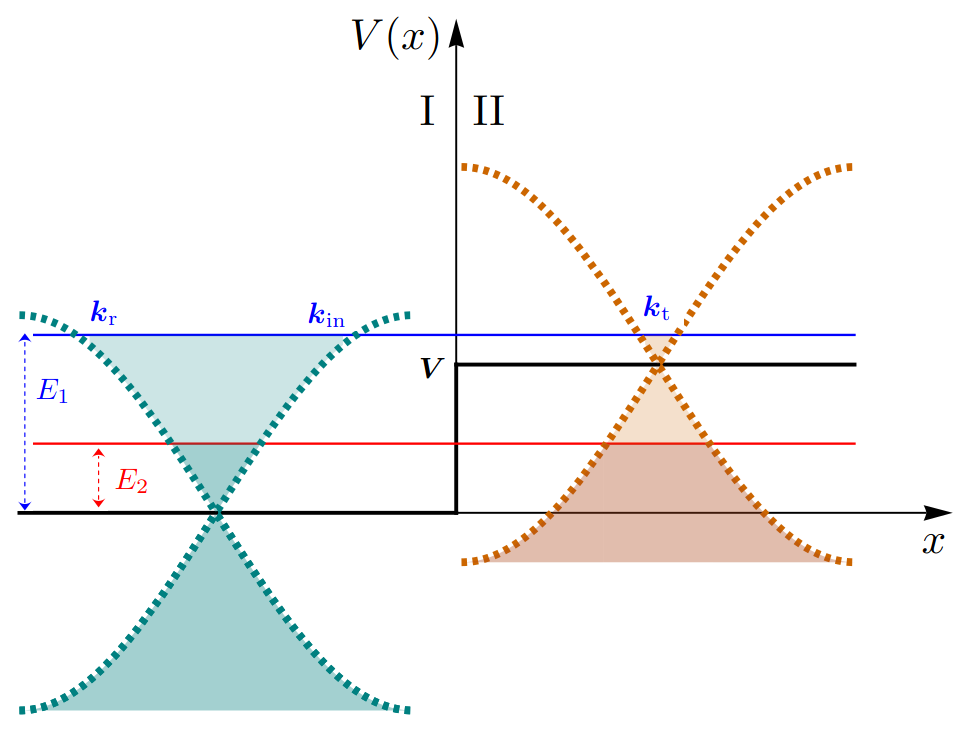}
    \end{tabular}
    \begin{tabular}{cc}
    (b) \qquad \qquad \qquad \qquad \qquad \qquad \qquad \qquad & \\
    \includegraphics[width=0.87\linewidth]{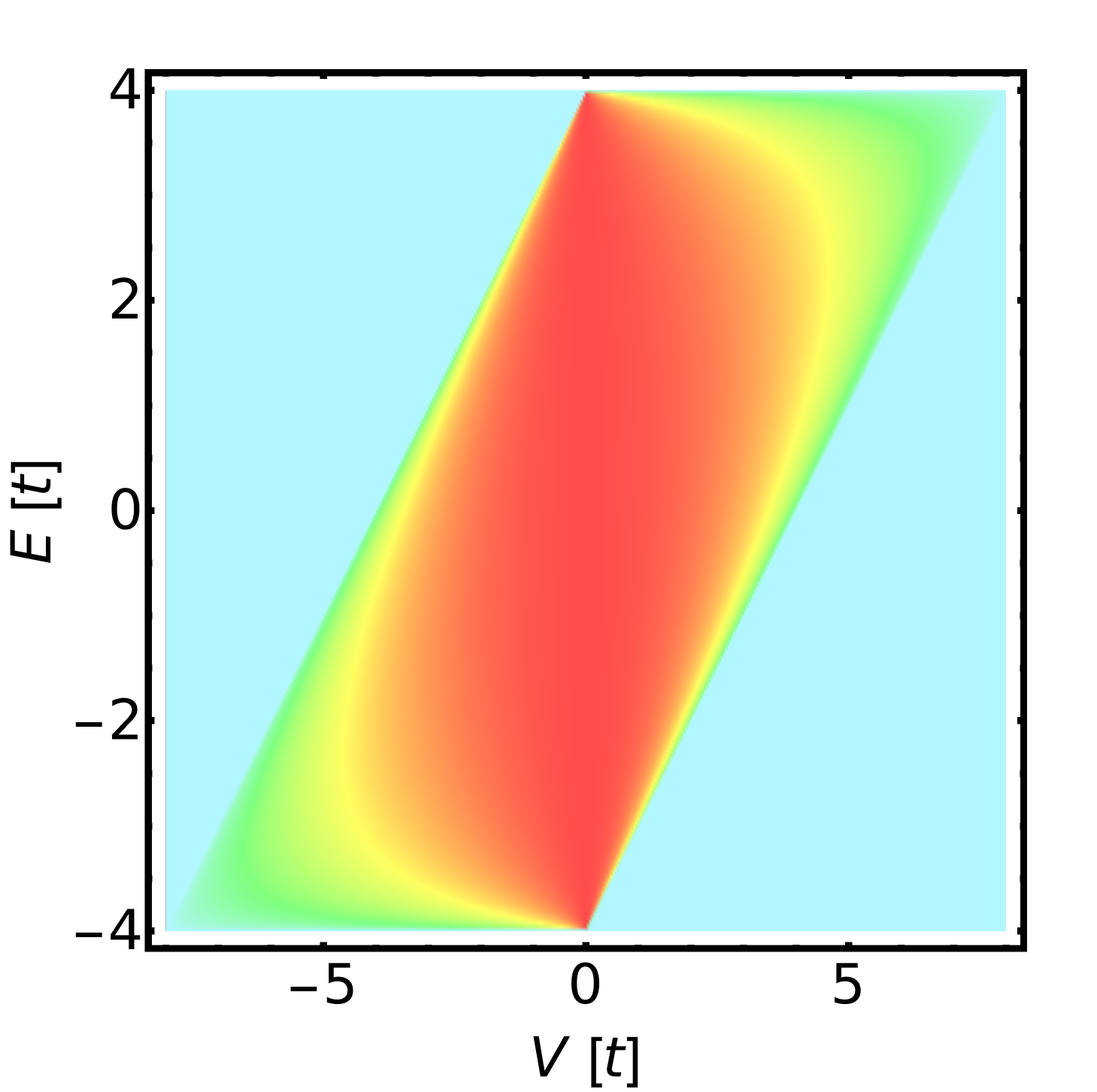} & 
    \includegraphics[width=0.1\linewidth]{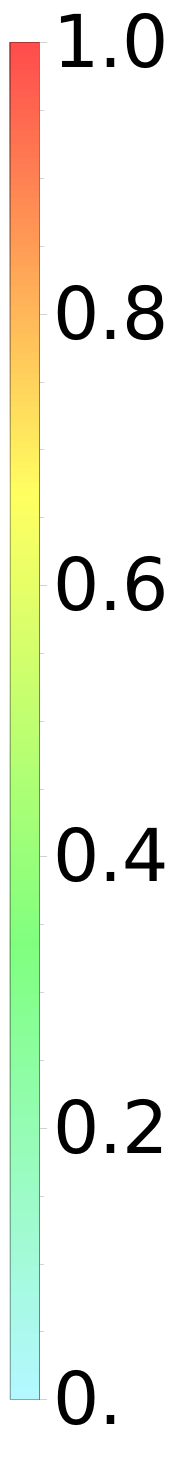}
    \end{tabular}
    \caption{(a) Electronic band structure of a pn junction (dashed green and orange curves)
based on the SSH lattice. The potential step $V(x)$ shifts the energy bands in region II (orange dotted curves) relative to region I (green dotted curves) by an energy $V$. The energy bands depend on the wave number and not the position. For illustrative purposes, we have drawn the shape of the dispersion relation in the potential $V(x)$. The horizontal blue and red lines represent two different energy levels, $E_1$ and $E_2$, respectively. At $E_1$, the blue line illustrates a transmission process within the conduction bands, with the incident ($k_\textrm{in}$), reflected ($k_\textrm{r}$), and transmitted ($k_\textrm{t}$) wave numbers indicated. (b) Transmission coefficient as a function of the Fermi level $E$ and potential $V$ for the $pn$ junction. Reddish (Bluish) region correspond to high (low) transmission.}
    \label{pnjunc}
\end{figure}

These matching conditions \eqref{TCC} are applied to the wave functions 
\begin{equation}\label{PsiI}
    \vec{\Psi}_\textrm{I}(x) = \vec{u}(k_\textrm{in})\textrm{e}^{ik_\textrm{in}x} + {A_\textrm{r}}\vec{u}(k_\textrm{r})\textrm{e}^{ik_\textrm{r}x}, 
\end{equation}
\noindent in region I, and

\begin{equation}\label{PsiII}
    \vec{\Psi}_\textrm{II}(x) = A_\textrm{t}\vec{u}(k_\textrm{t})\textrm{e}^{ik_\textrm{t}x} 
\end{equation}

\noindent in region II for the $pn$ junction shown in Fig. \ref{pnjunc}(a). The operators $\hat{Q}_j$ work just in the plane wave parts

\begin{eqnarray}
    \hat{Q}_j\textrm{e}^{ik x} & = & \textrm{e}^{ik x}\sum^\infty_{n = 0}\frac{(i k)^{(j+1)n}}{(n+1)!} \nonumber\\
    & = & \textrm{e}^{ik x}\left(\frac{\textrm{e}^{ik(-1)^{j+1}} -1}{k}\right),
\end{eqnarray}

\noindent where $k = k_\textrm{in/r/t}$. By using these relations for the wave functions \eqref{PsiI} and \eqref{PsiII}, and arranging them in a matrix form, we have

\begin{equation}\label{AryAt}
    \vec{w}(k_\textrm{in}) + A_\textrm{r}\vec{w}(k_\textrm{r}) = A_\textrm{t}\vec{w}(k_\textrm{t}).
\end{equation}

\noindent The quantities $\vec{w}(k)$ are defined by

\begin{align}\label{wtild}
   \vec{w}(k)  = (f(k)u^{(1)}(k),f^*(k)u^{(2)}(k)),
\end{align}

\noindent where the functions $f(k)$ are

\begin{equation}
   f(k) = \frac{\textrm{e}^{i k} -1}{k}.
\end{equation}

\noindent The conservation energy $E$ allows us to relate $k$ in terms of $E$, the barrier height $V$, and hopping parameters

\begin{eqnarray}\label{ks}
    k_\textrm{in}(E) & = & -k_\textrm{r}(E) \nonumber\\
    & = & \sgn(E)\arccos\left(\frac{t^2+t'^2-E^2}{2\,t\,t'}\right)\nonumber\\
    k_\textrm{t}(E,V)  & = & \sgn(E-V)\arccos\left(\frac{t^2+t'^2-(E-V)^2}{2\,t\,t'}\right).\nonumber\\
    &&
\end{eqnarray}

\noindent With these relations in Eq. \eqref{ks}, the reflection and transmission coefficients are given by 

\begin{equation}\label{transm}
    T=1-R = 1-\frac{|\vec{w}(k_\textrm{in})\times\vec{w}(k_\textrm{t})|^2}{|\vec{w}(k_\textrm{r})\times\vec{w}(k_\textrm{t})|^2},
\end{equation}

\noindent which can be expressed as a function of $E$ and $V$. The transmission features, as shown in Fig. \ref{pnjunc}(b), are explained by using the relative pseudo-spin direction \cite{BetancurOcampo2026}

\begin{equation}
    \gamma(k_\textrm{in},k_\textrm{t})=\arccos\left(\frac{|\vec{w}^*(k_\textrm{in})\cdot\vec{w}(k_\textrm{t})|}{|\vec{w}(k_\textrm{in})||\vec{w}(k_\textrm{t})|}\right).
\end{equation}

The transmission probability is highly angle-dependent. For propagating states, the angles $\gamma = 0$ and $\pi/2$ denote the regimes of Klein and anti-Klein tunneling, respectively \cite{BetancurOcampo2026}. Fig. \ref{pnjunc}(b) displays the transmission from Eq. \eqref{transm} as a function of the Fermi level $E$ and the potential $V$ for a symmetric SSH-lattice $pn$ junction ($t = t'$). The reddish region highlights high transmission, which occurs at lower potential values where electron propagation is nearly free. Conversely, the bluish region indicates internal reflection associated with evanescent modes. This transmission profile serves as a baseline reference for analyzing how a Wannier-like potential modifies transport in the subsequent sections.

\section{Global matrix framework}
\begin{figure*}
    \centering
    \includegraphics[width=1\linewidth]{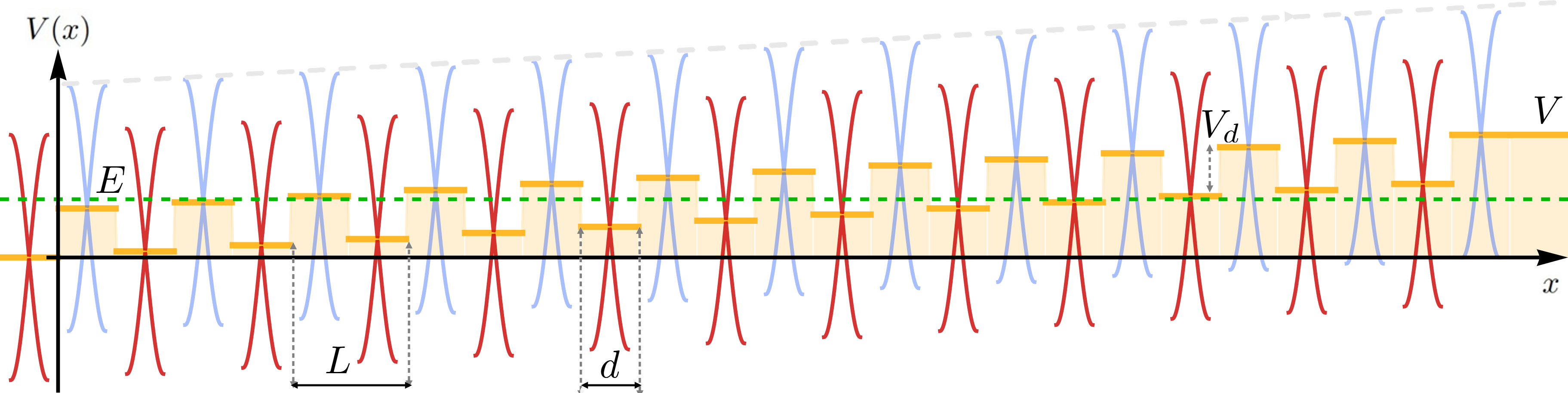}
    \caption{Schematic diagram of the local electronic band structure in the SSH lattice under a stratified potential mimicking a Wannier-Stark profile. The potential step structure (orange blocks) features a spatial period $L$, well width $d$, and well depth $V_d$. Red and light-blue curves illustrate the local valence and conduction bands of the SSH lattice, which shift upward along the chain. The dashed green line indicates the Fermi level $E$, and $V$ denotes the constant potential in the outgoing region. The set of parameters are $d = 10$ (in units of the lattice constant $a = 1$), $L = 20$, electron charge $e = 1$, the electric field $F = 0.01$, the Fermi level $E = 1.9$, and $V_0 = 1.4$, and hopping parameters $t = t' = 2$. The total number of barriers plus wells is $N = 25$. The gray dashed line is drawn to guide the eye. It has a slope equal to the electric field and touches the border of the blue energy bands.}
    \label{fig:Wanpot}
\end{figure*}

 To study electron transmission under stratified electrostatic potentials, such as the Wannier-like potential shown in Fig. \ref{fig:Wanpot}, we partition the system into a sequence of adjacent potential barriers. Each barrier $n$ is characterized by a width $x_n - x_{n-1}$ and a constant height $V_n$. Within this framework, the Global Matrix Method (GMM) begins by defining the wavefunction ansatz in the $n$-th barrier as

\begin{equation}\label{WFbarr}
    \vec{\Psi}^n(x) = A_{n}\vec{u}(k^a_n)\textrm{e}^{ik^a_n x}+ B_{n}\vec{u}(k^b_n)\textrm{e}^{ik^b_n x},
\end{equation}

\noindent where $A_{n}$ and $B_{n}$ are the amplitudes of the wave inside the barrier. The wave numbers inside the barrier $n$ are 
\begin{eqnarray}\label{kns}
    k^a_n & = & -k^b_n \nonumber\\
    & = & \sgn(E-V_n)\arccos\left(\frac{t^2+t'^2-(E-V_n)^2}{2\,t\,t'}\right).\nonumber\\
    & &
\end{eqnarray}

For the first region, $n = 0$, the incident wave has an amplitude $A_0 = 1$ and wave number $k^a_0=k_\textrm{in}$, for the reflected part, the amplitude is $B_0 = A_r$ and wave number $k^b_0=k_\textrm{r}$. With $n = N+1$ in the last region, $B_{N+1} = 0$ because there is no reflection, and the transmitted amplitude is $A_{N+1} = A_t$.

Applying the matching conditions of Eq. \eqref{TCC} at $x = x_{n}$ that separates the regions with wave functions $\vec{\Psi}^n(x)$ and $\vec{\Psi}^{n+1}(x)$, we have

\begin{equation}
    M_{n+1 n}\left(\begin{array}{c}
          A_{n+1}\\
          B_{n+1}
    \end{array}\right) = M_{n n}\left(\begin{array}{c}
          A_{n}\\
          B_{n}
    \end{array}\right),
\end{equation}

\noindent where the matrices $M_{m l}$ are

\begin{equation}\label{Ms}
    M_{m l} =\left[\begin{array}{cc}
      \vec{w}(k^a_n)  & \vec{w}(k^b_n)
    \end{array}\right]\left(\begin{array}{cc}
      \textrm{e}^{ik^a_{n}x_l}  & 0\\
      0 & \textrm{e}^{ik^b_{n}x_l}
    \end{array}\right),
\end{equation}

\noindent which taken into account the substitution $\vec{u}(k_n)\rightarrow\vec{w}(k_n)$ given by Eq. \eqref{wtild}. It is possible to get a global equation system 
\begin{equation}\label{Gsys}
    G\vec{X}=\vec{b}
\end{equation}

\noindent involving all the coefficients $A_n$ and $B_n$ by defining the square global matrix 

\begin{align}
\label{eq:Nssh}
    G = 
        \left(\begin{array}{c c c c c}
        \vec{w}(k_\textrm{r})  & -M_{10} & 0_{2\times2} & \cdots & 0_{2\times1}  \\
         0_{2\times1} & M_{11} & -M_{21}  & 0_{2\times2} & \cdot\\
         0_{2\times1} & 0_{2\times2} & \cdot & \cdot & \cdot \\
         \vdots & \vdots &  \vdots & \vdots & 0_{2\times1} \\
         0_{2\times1} & 0_{2\times2} & 0_{2\times2} & M_{NN} & -\vec{w}(k_\textrm{t})
    \end{array}\right)
\end{align}

\begin{figure*}
    \centering
    \begin{tabular}{cccc}
    (a)\quad \qquad \qquad $N = 3$  \quad \qquad \qquad & (b)\quad \qquad \qquad $N = 11$ \quad \qquad \qquad & (c)\quad \qquad \qquad $N = 25$ \quad \qquad \qquad\\
        \includegraphics[width=0.3\linewidth]{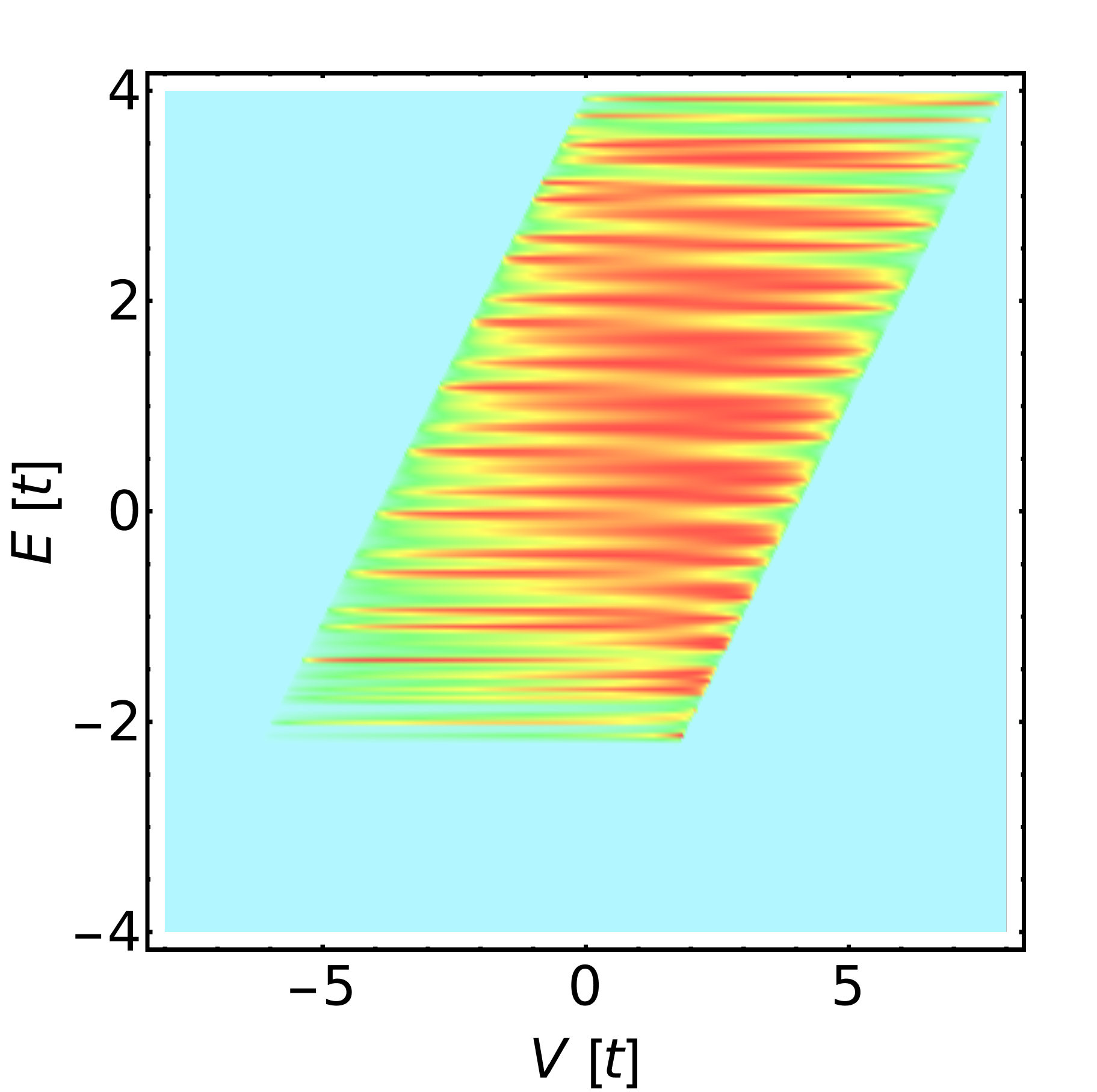} &  
        \includegraphics[width=0.3\linewidth]{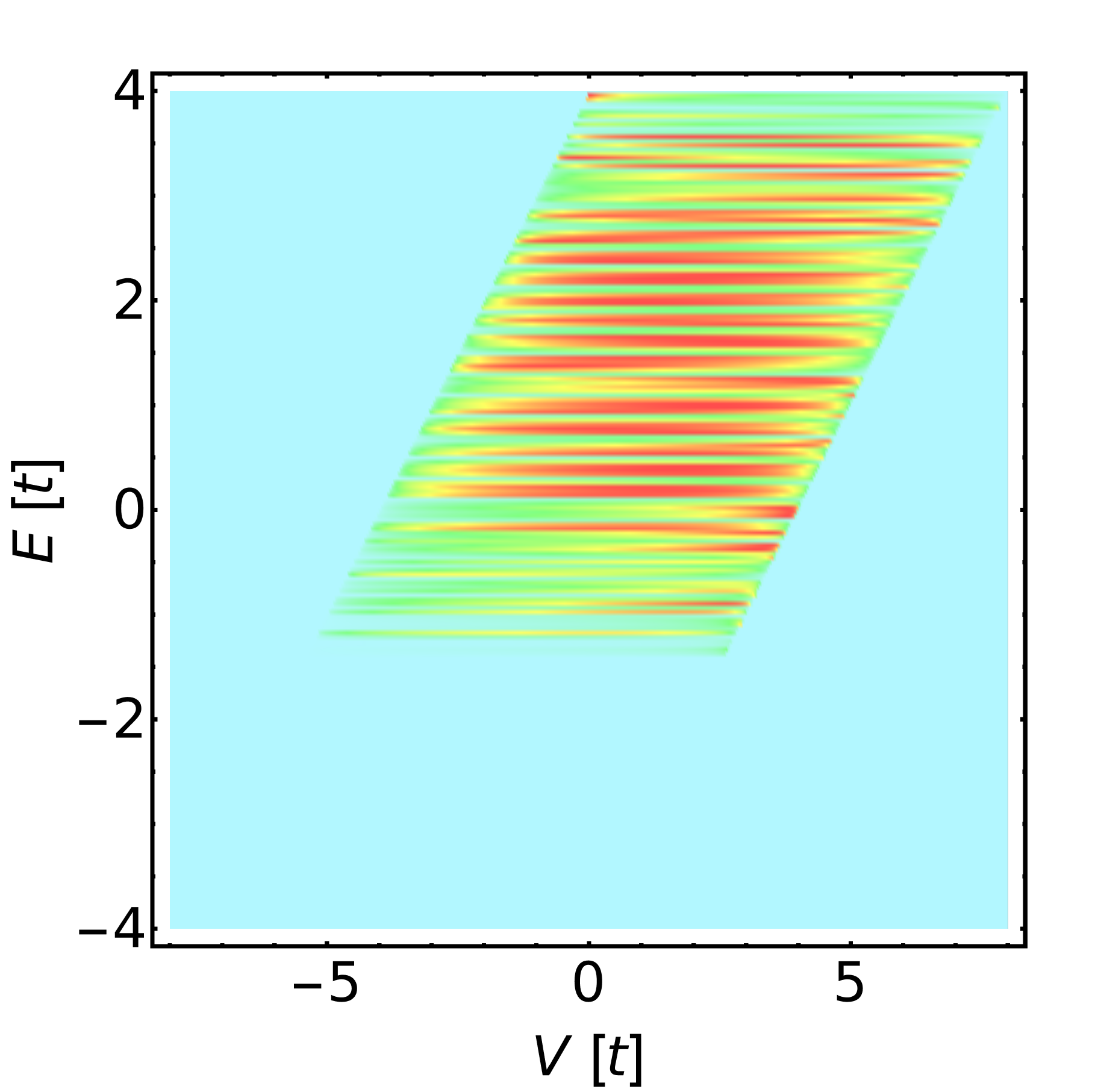}&
        \includegraphics[width=0.3\linewidth]{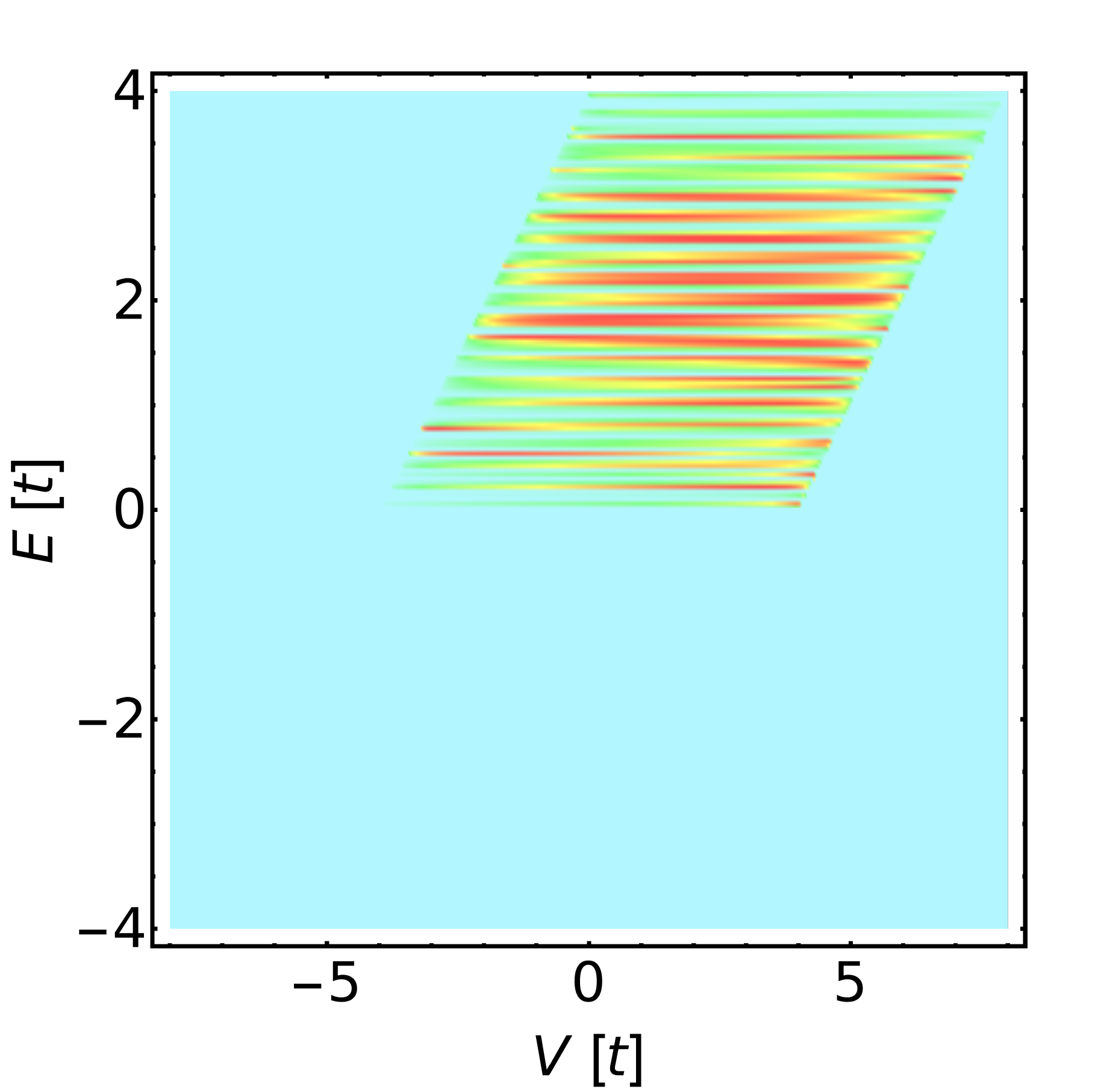}&
        \includegraphics[width=0.035\linewidth]{Barra.pdf}
    \end{tabular}
    \caption{Electron transmission through a Su-Schriefer-Heeger lattice in the presence of a stratified Wannier-like potential, shown as a function of the Fermi energy $E$ and the potential in the last region $V$, for (a) $N = 3$, (b) $N = 11$, and (c) $N = 25$. The set of parameters values of the tight-binding and Wannier-like potential are the same than in Fig. \ref{fig:Wanpot}.}
    \label{fig:Transm_Wannier}
\end{figure*}

\noindent which is a matrix of dimension $2(N+1)\times2(N+1)$, where the vector $\vec{X} \equiv (A_\textrm{r},\dots,A_n,B_n,\ldots,A_\textrm{t})$ contents the wave amplitudes and $\vec{b} = (-\vec{w}(k_\textrm{in}),0,\ldots,0)$.

\noindent The square matrix $G$ contains the complete information regarding electron transport through the stratified electrostatic potential. This method has consistently reproduced established transport behaviors in systems such as graphene and one-dimensional chains ~\cite{Allain2011,Katsnelson2006,BetancurOcampo2026}. The main advantage of the GMM is that it solves for the wavefunction amplitudes, $A_n$ and $B_n$, simultaneously across the entire system. Crucially, although the local wave function ansatz in Eq. \eqref{WFbarr} is defined piecewise, solving the linear system in Eq. \eqref{Gsys} alongside the mapping $\vec{u}(k)\rightarrow \vec{w}(k)$ via the boundary conditions ensures a continuous global wave function across the entire chain

\begin{equation}\label{Phi}
    \vec{\Phi}_n(k) = A_{n}\vec{w}(k^a_n)\textrm{e}^{ik^a_n x}+ B_{n}\vec{w}(k^b_n)\textrm{e}^{ik^b_n x}.
\end{equation}

\noindent To analyze the change of the relative pseudospin $\gamma$ for stratified potential media, it is necessary to calculate the transfer matrix \cite{BetancurOcampo2026,DiazBautista2024,MarinColli2026,Jellal2025,IbarraReyes2025}

\begin{equation}\label{Lambda}
    \Lambda = \prod^N_{n=1}M_{nn-1}M^{-1}_{nn},
\end{equation}

\noindent from the matrices defined in Eq. \eqref{Ms} and the relative pseudospin is given by

\begin{equation}\label{gamma}
     \gamma=\arccos\left(\frac{|\vec{w}^*(k_\textrm{in})\cdot\Lambda\vec{w}(k_\textrm{t})|}{|\vec{w}(k_\textrm{in})||\Lambda\vec{w}(k_\textrm{t})|}\right).
\end{equation}

We employ the GMM to model wave transmission across the Wannier potential illustrated in Fig. \ref{fig:Wanpot}. The incident region ($x < 0$) is set to a constant potential $V(x) = 0$. Within the scattering region ($0 \leq x \leq NL$), the Wannier-like potential is given by

\begin{align}\label{Vw}
    V_\textrm{W}(x) = V_d\sum^\infty_{n=0}[\Theta(x-nL)-\Theta(x-nd)] + & \nonumber\\
    eFL\sum^\infty_{n=0}\Theta(x-nL),& 
\end{align}

\noindent which describes a stratified medium composed of successive potential barriers and wells whose heights and depths increase linearly. Here, $\Theta(x)$ is the Heaviside step function, $V_d$ represents the well depth, $d$ is the well width (see Fig. \ref{fig:Wanpot}), $L$ is the spatial period, and $F$ denotes the electric field strength. The transmission region ($x > NL$) is maintained at a tunable, constant potential $V$.

In the presence of a uniform electric field and a periodic potential $V_p(x)$ of period $L$, the standard Schr\"odinger Hamiltonian $H_\textrm{Sch}=p^2/2m + V_p(x)+eFx$ yields the characteristic Wannier-Stark ladder spectrum, $E_n = E+neFL$ \cite{Wannier1960,Hartmann2004}. Similarly, the SSH Hamiltonian in Eq. \eqref{Hssh} subjected to this biased potential, $H' = H_\textrm{SSH}+ V_p(x)+eFx$, also preserves the validity of this spectral structure. In the following section, we demonstrate that these Wannier-Stark ladders manifest as antiresonances under the potential in Eq. \eqref{Vw}, with a energy spacing that matches the ladder separation $\Delta E = eFL$.

\section{Signature of Wannier-Stark ladder from anti-resonances}
We calculated the transmission for three Wannier potential structures, with $N = 3, 11$, and $25$, as a function of the Fermi level and the terminal region potential $V$, using the parameters $t = t' = 2$, $e = 1$, $d = 10$, $L = 20$, and $F=0.01$, see Fig. \ref{fig:Transm_Wannier}. Unlike the $pn$ junction in Fig. \ref{pnjunc}(b), the alternating barriers and wells of the Wannier-like potential generate an interference pattern. When incident waves impinge on the potential for lower negative energies from the valence band, total internal reflection emerges due to evanescent modes by the uniform electric field $eFx$. However, in the energy range hosting continuous propagating modes, as shown in Fig. \ref{fig:Wanpot}, a sequence of $V$-independent total reflections emerges equally spaced. Such antiresonances are more evident for larger values of $N$ (see Fig. \ref{fig:Transm_Wannier}(c)), and match with the Wannier-Stark ladder energy spacing ($FL = 0.2$). We term these features Wannier-Stark antiresonances, which crucially differ from conventional Wannier-Stark ladders by the absence of Bloch oscillations \cite{Bloch1929,Wannier1960,Wannier1970,Hartmann2004}.

\begin{figure*}
\begin{tabular}{ccc}
    \centering
    (a)\quad \qquad \qquad $N = 3$  \quad \qquad \qquad & (b)\quad \qquad \qquad $N = 11$ \quad \qquad \qquad & (c)\quad \qquad \qquad $N = 25$ \quad \qquad \qquad\\
    \includegraphics[width=0.33\linewidth]{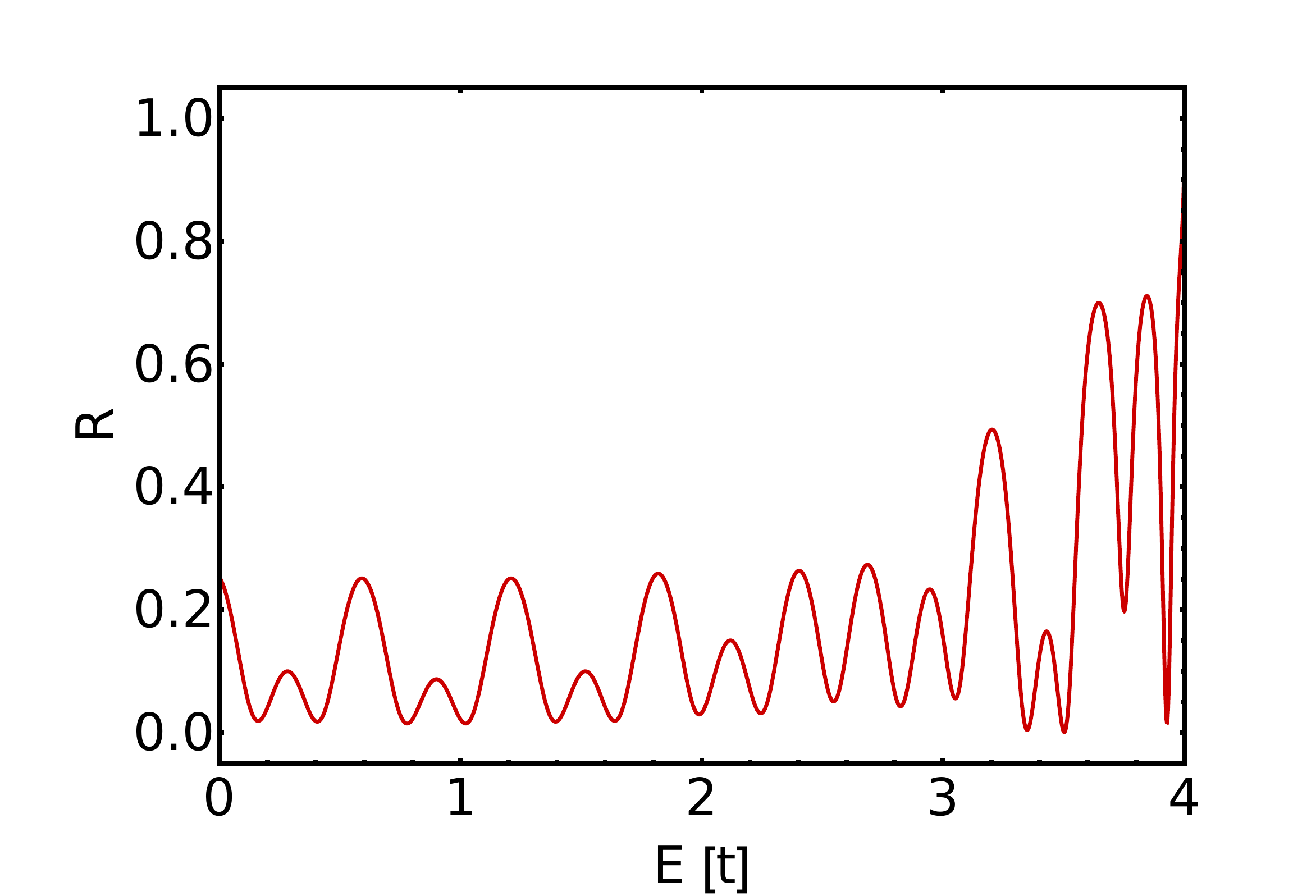}&
    \includegraphics[width=0.33\linewidth]{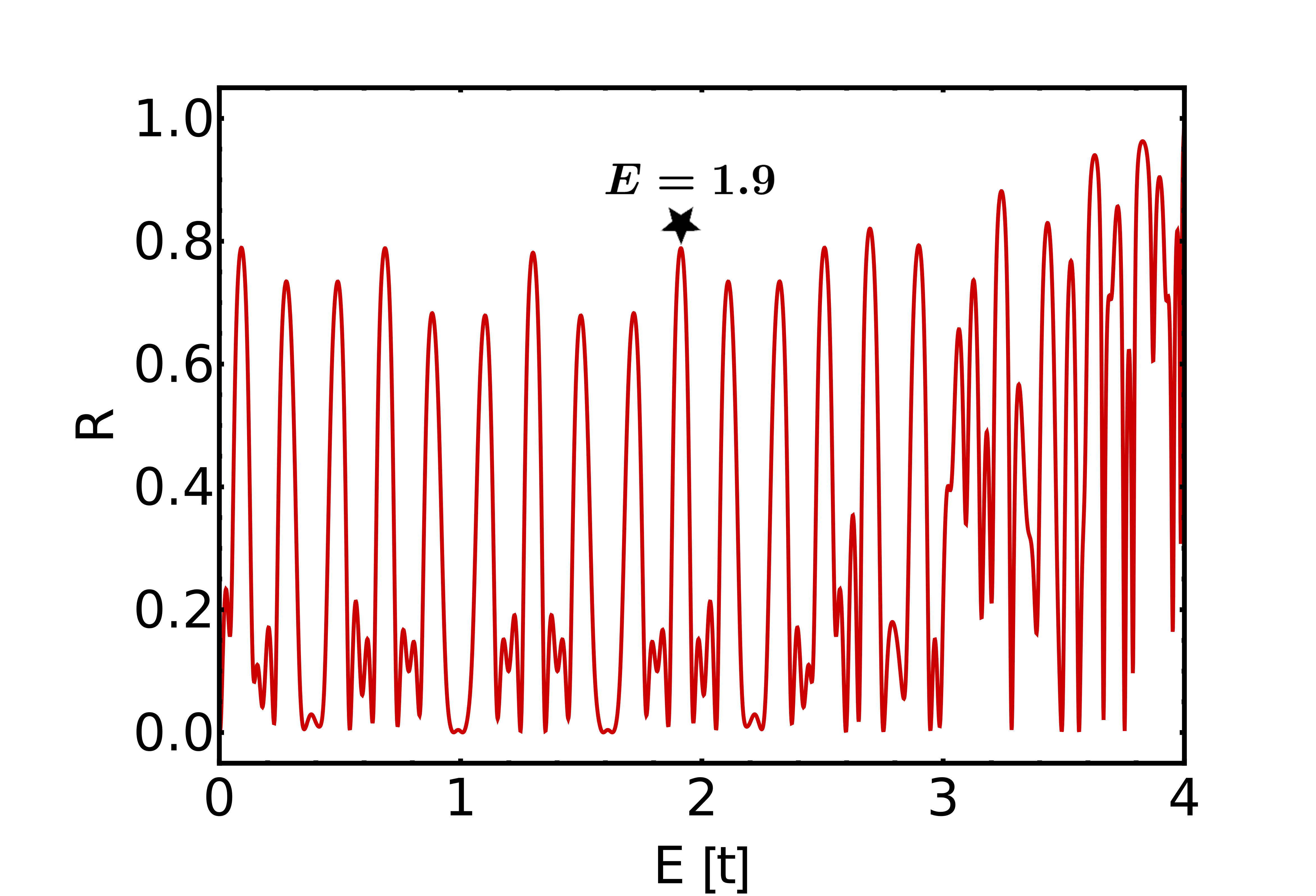}&
    \includegraphics[width=0.33\linewidth]{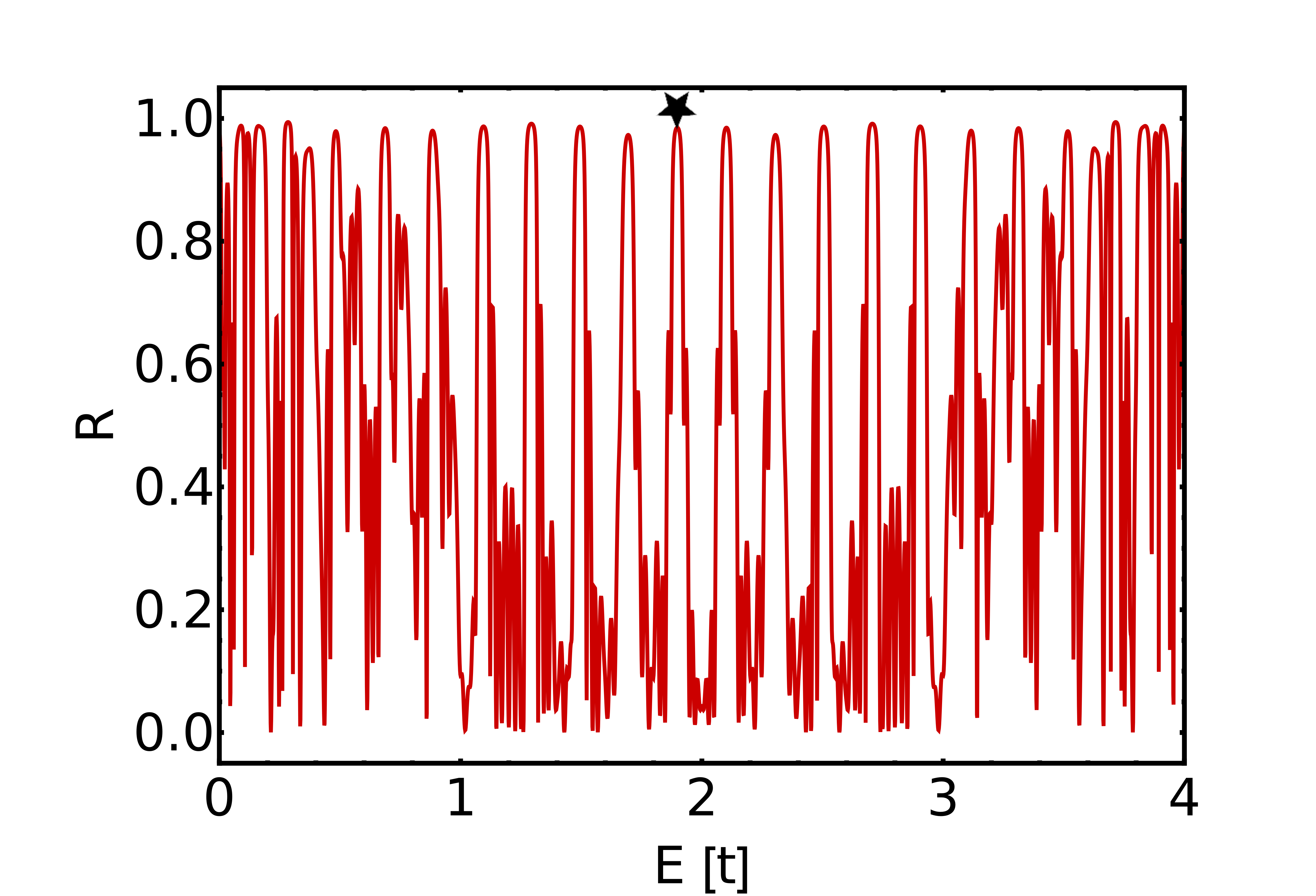}\\
    (d)\quad \qquad \qquad \qquad  \quad \qquad \qquad & (e)\quad \qquad \qquad \qquad  \quad \qquad \qquad & (f)\quad \qquad \qquad \qquad  \quad \qquad \qquad\\
    \includegraphics[width=0.33\linewidth]{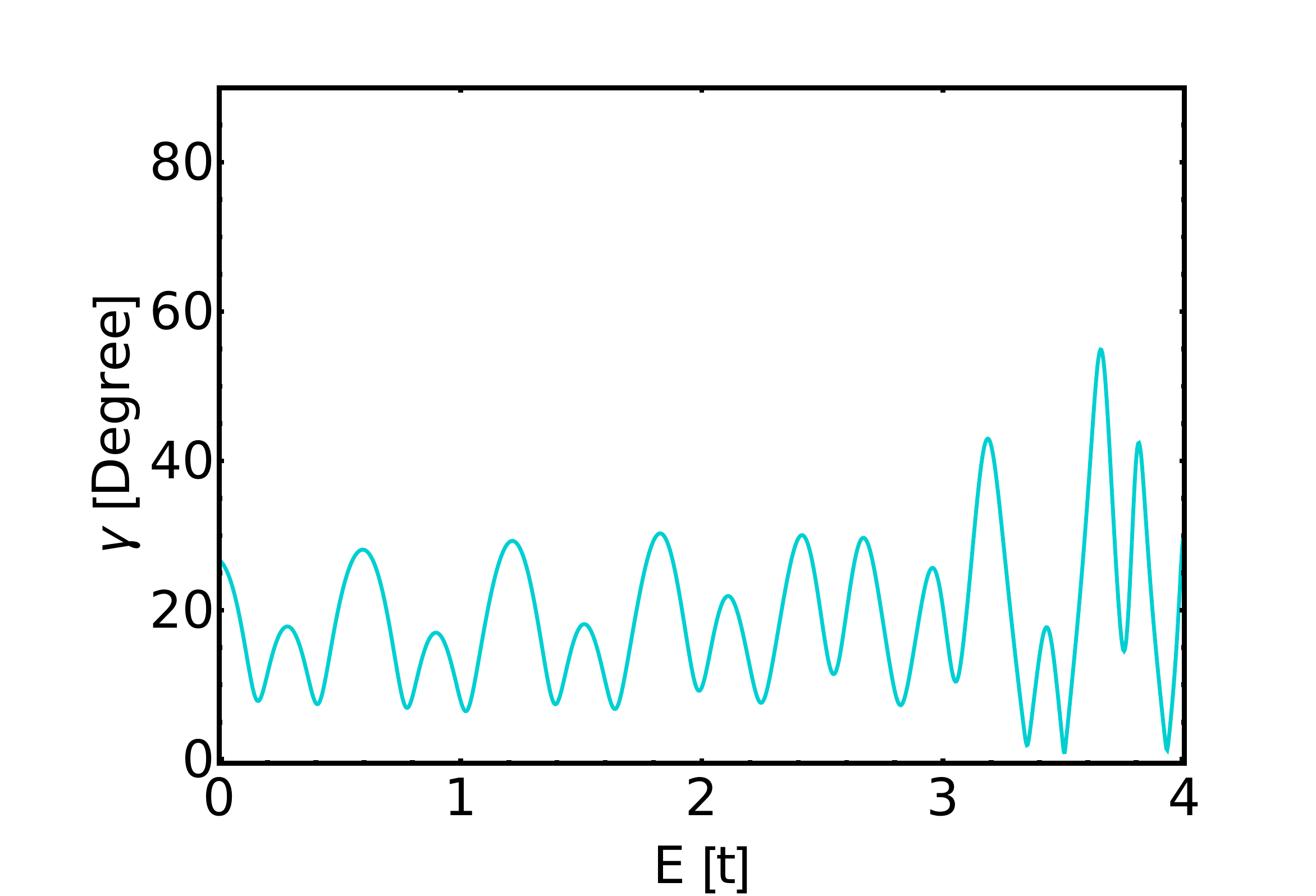}&
    \includegraphics[width=0.33\linewidth]{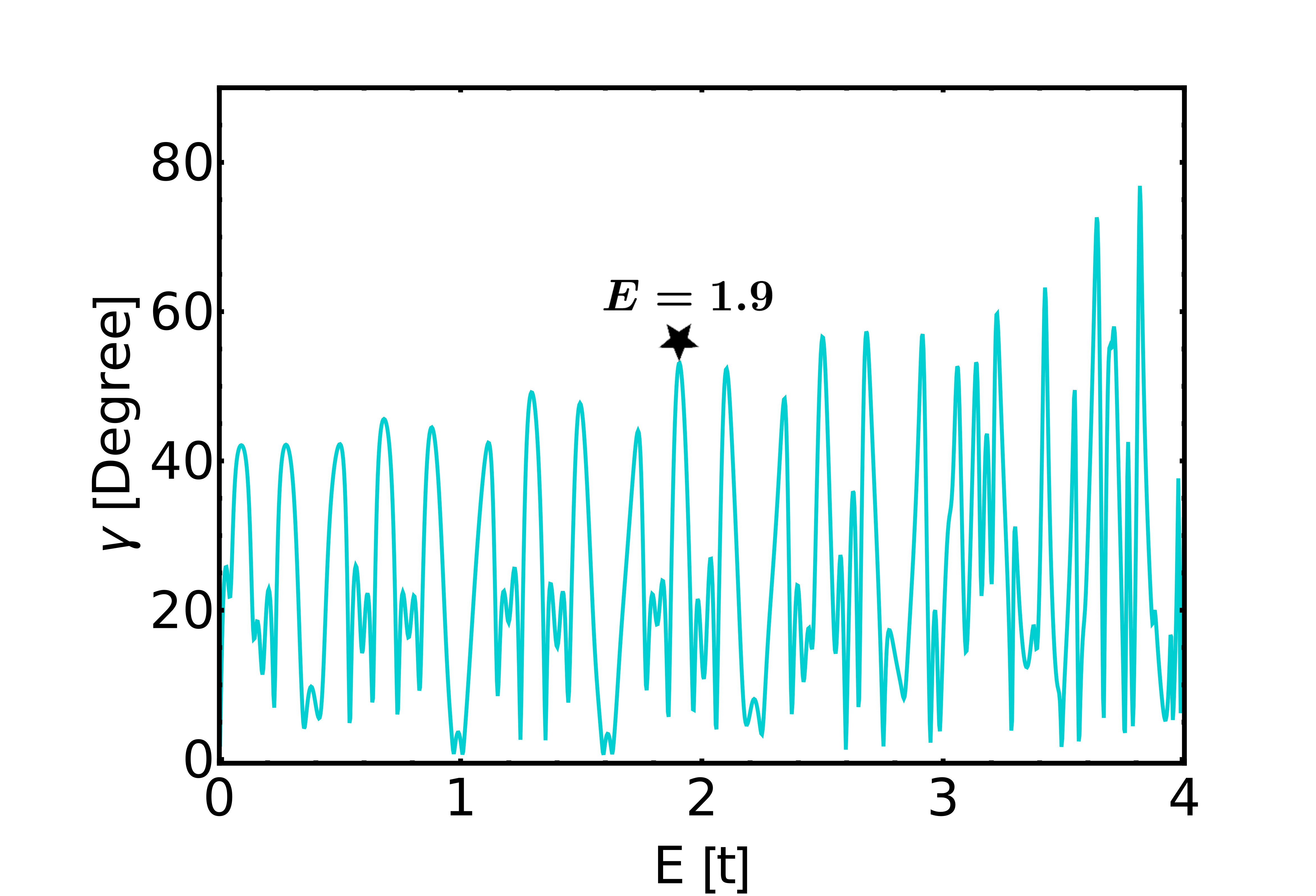}&
    \includegraphics[width=0.33\linewidth]{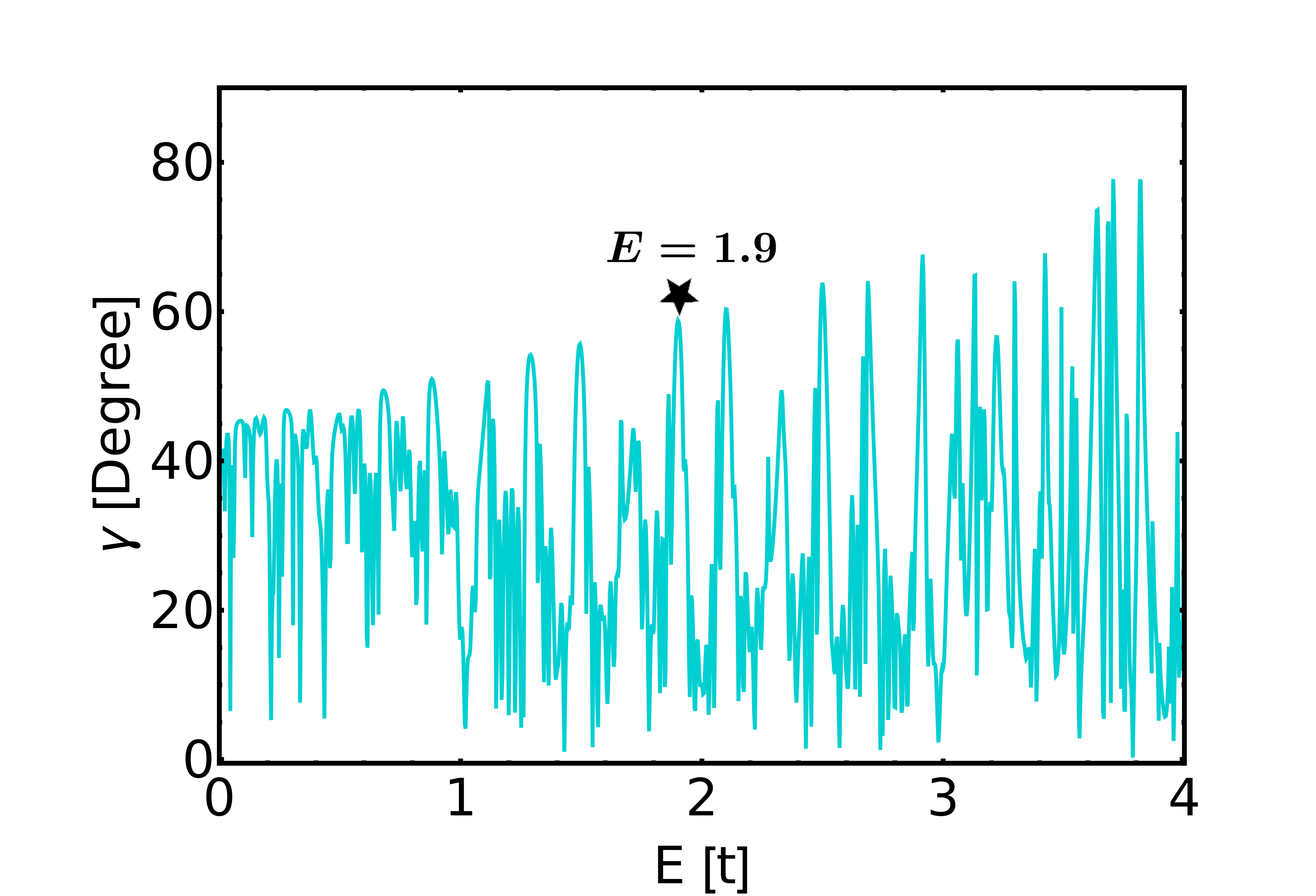}
    \end{tabular}
    \begin{tabular}{cc}
    (g)\quad \qquad \qquad \qquad  \quad \qquad \qquad& (h)\quad \qquad \qquad \qquad \quad \qquad \qquad\\
    \includegraphics[width=0.315\linewidth]{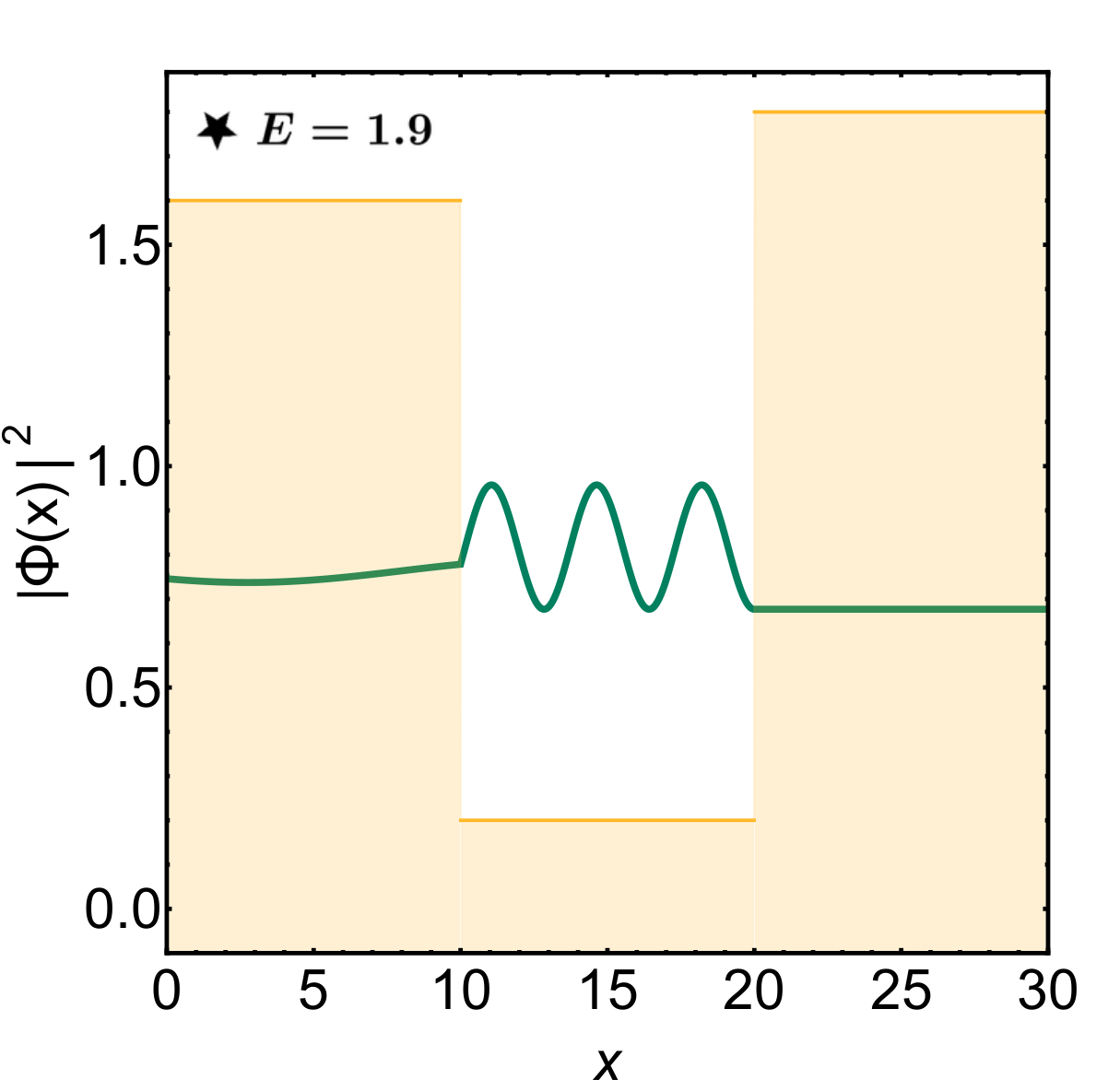}&
    \includegraphics[width=0.315\linewidth]{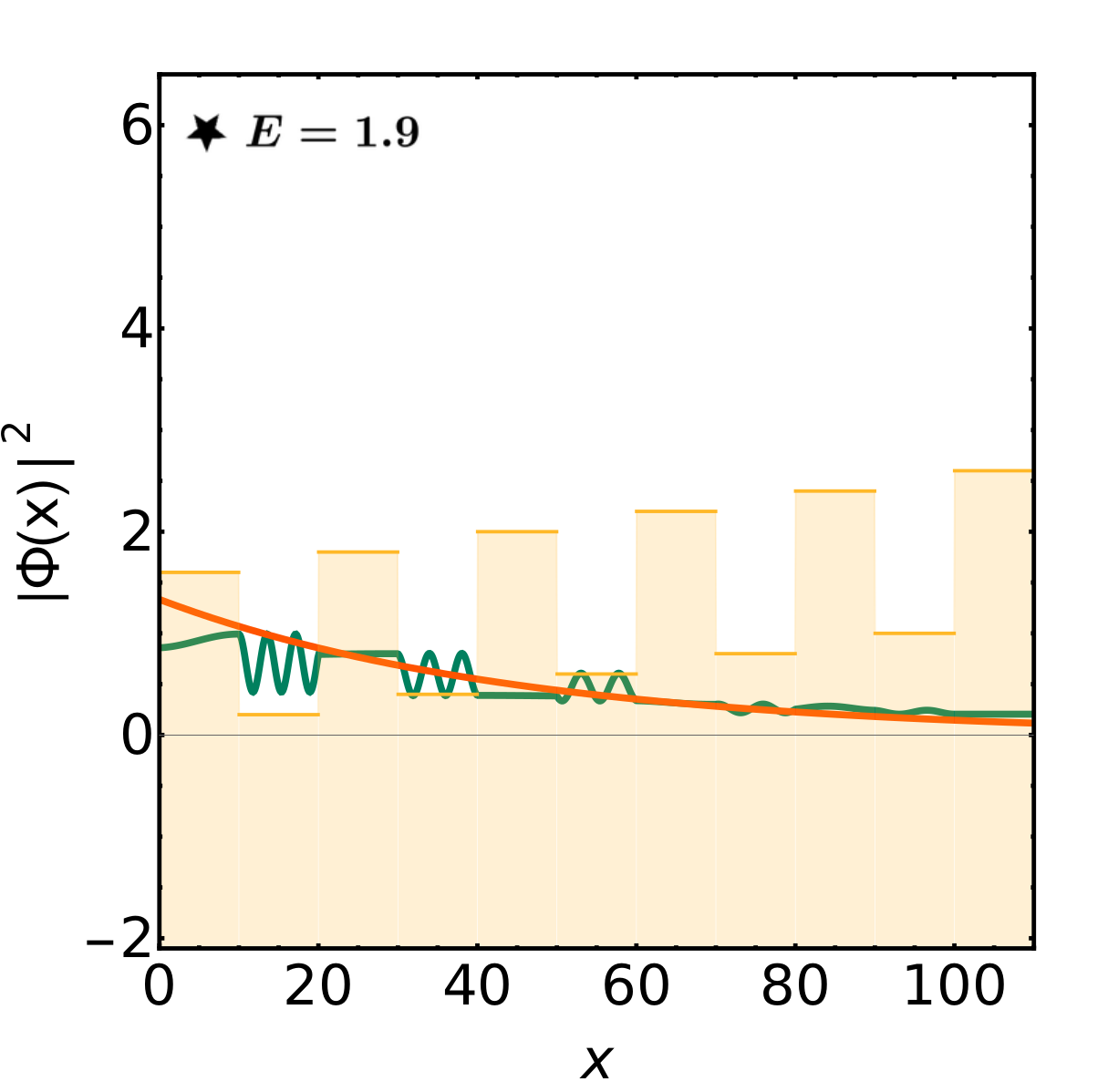}
    \end{tabular}
    \begin{tabular}{c}
    (i)\quad \qquad \qquad \qquad \qquad \quad \qquad \qquad \qquad \qquad \qquad \qquad \qquad \qquad \qquad \qquad\\
     \includegraphics[width=0.68\linewidth]{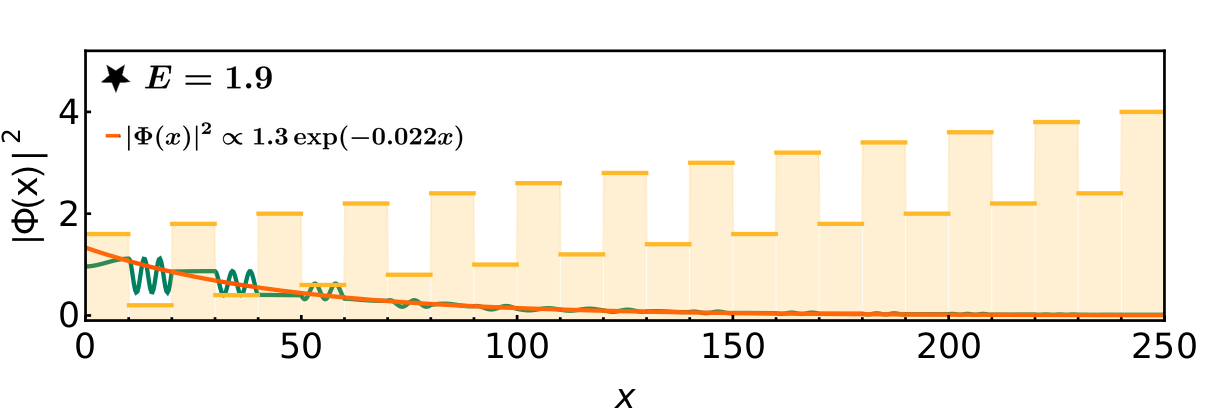}
    \end{tabular}
    \caption{(a)-(c) Reflection coefficient as a function of the energy $E$ for $N = 3, 11$, and $25$, the antiresonances are evidenced and their relative energy separation is $FL = 0.2$. (d)-(e) Relative pseudospin as a function of the energy $E$, obtained from Eq. \eqref{gamma}, for the reflections presented in (a)-(c). (g)-(i) Squared wave function (Eq. \eqref{Phi}) calculated using the matching conditions across the Wannier-like potential for the structure depicted in (a)-(c). The set of parameters values of the tight-binding and Wannier-like potential are the same than in Fig. \ref{fig:Wanpot}. The black star indicates one of the twenty antiresonances at energy $E = 1.9$.}
    \label{fig:AWS}
\end{figure*}

To clearly visualize these antiresonances, we show the reflection as a function of energy in Fig. \ref{fig:AWS}(a)-(c), where the equally spaced peaks become apparent for $N \geq 11$. We then computed the relative pseudo-spin angle to determine the origin of these reflections, employing the matrix transfer method from Eqs. \eqref{Lambda} and \eqref{gamma}, as shown in Fig. \ref{fig:AWS}(d)-(f). It is important to note that the Wannier-Stark antiresonances are directly associated with the maxima of the relative pseudo-spin angle $\gamma$. As waves propagate across the potential profile, they undergo multiple reflections and may eventually return to the incident side. Nevertheless, Fabry-Pérot resonances arise from constructive interference, and therefore, electrons fully traverse the Wannier-like potential. These correspond to the minima in the relative pseudospin angle $\gamma$. 

To examine how the wave function associated to a specific Wannier-Stark antiresonance behaves across the potential, we show the squared modulus of the wave function in Eq. \eqref{Phi} throughout all barriers and wells as a function of position in Fig. \ref{fig:AWS}(g)-(h). We choose the energy $E = 1.9$, indicated by a black star in Fig. \ref{fig:AWS}, which is one of the twenty possible antiresonances in the energy range from 0 to $4$ occurs.  Note that this wave function is continuous because the matching conditions require substituting the eigenvector $\vec{u}(k)$ with the effective pseudo-spin vector $\vec{w}(k)$. For $N = 3$, shown in Fig. \ref{fig:AWS}(g), the wave function extends across the entire potential profile. In the wells, the wave function exhibits an oscillating pattern due to the reduced wavelength, whereas inside the barriers, the oscillations are damped because the wavelength is larger. In Fig. \ref{fig:Wanpot}, the Fermi level at $E = 1.9$ cuts across all the energy bands, which accounts for this alternation between shortened and enlarged wavelengths as the electron traverses barriers and wells. When $N$ is increased, the exponential decay is more evident, $|\Phi(x)|^2\propto |\Phi_0|^2\exp(-x/\lambda_0)$, with a best fit yielding $|\Phi_0|^2 \approx 1.3$ and $\lambda_0 \approx  45.1$. It is worth emphasizing that the observed localization associated with Wannier-Stark antiresonances arises in propagating modes, because the electron always passes from one dispersive band to another without encountering a bandgap, as shown in Fig. \ref{fig:Wanpot}.  

\section{Conclusions and final remarks}
We studied the wave transmission in a one-dimensional Su-Schrieffer-Heeger lattice in the presence of a Wannier-like potential consisting of a periodic and uniform electric field potential. By employing the global matrix method and wave function matching conditions within the full band of the tight-binding model, we successfully identified the emergence of Wannier-Stark antiresonances. These antiresonances are equally spaced in energy, and their separation obeys the Wannier-Stark ladder spectrum. Moreover, the amplitude of the wave crossing the potential has an exponential decay for propagation modes. Consequently, these antiresonances offer a distinct experimental signature that can be recognized through reflectance measurements and wave function localization, with their sharpness becoming more pronounced as the system size increases.  An interesting outlook for future study is exploring the interplay between these antiresonances and the topologically protected edge states of the SSH lattice. While this work is originally formulated for electron scattering in Su-Schrieffer-Heeger chains, the underlying wave mechanics can be feasibility extended to other classical analogues. Specifically, in platforms like acoustic metamaterials, photonic lattices, or water waves, the required linear potential gradient can be easily engineered, opening avenues for non-electronic wave manipulation and filtering devices.  

\bibliography{Wannier}
\end{document}